\documentclass[amsmath,amsfonts,aps,pra,%
superscriptaddress,%
reprint%
]{revtex4-2}
\usepackage[utf8]{inputenc}
\usepackage[english]{babel}
\usepackage{color} 
\usepackage{graphicx}
\usepackage{mathrsfs} 
\usepackage{calc} 
\usepackage{array} 
\usepackage{titlesec} 
\usepackage{enumitem} 
\usepackage{hyperref} 
\usepackage{soul} 
\usepackage{setspace} 
\usepackage{lipsum}
\usepackage{float} 
\usepackage{cleveref} 
\usepackage{placeins} 
\usepackage[version=4]{mhchem}

\newcommand\mytitle{phaser: A unified and extensible framework for fast electron ptychography}

\hypersetup{
    pdftitle={\mytitle},
    pdfauthor={Colin Gilgenbach},
    pdfpagemode=FullScreen,
    pdfborder={0 0 0},
    colorlinks=false
}
\renewcommand{\AA}[0]{\text{\normalfont\r{A}}}
\newcommand{\code}[1]{\texttt{#1}}

\titlespacing*{\section}
{0pt}{12pt}{0pt}
\titlespacing*{\subsection}
{0pt}{8pt}{0pt}
\titlespacing*{\subsubsection}
{0pt}{8pt}{0pt}

\setlist[description]{leftmargin=2em, itemsep=0pt, parsep=12pt}

\begin{document}

\title[phaser: A unified framework]{\mytitle}
\author{Colin Gilgenbach}

\author{Menglin Zhu}

\author{James M.~LeBeau}
\email{lebeau@mit.edu}\affiliation{Department of Materials Science \& Engineering, Massachusetts Institute of Technology, Cambridge, MA 02472}
\date{\today}

\begin{abstract}
We present \code{phaser}, an open-source Python package that provides a unified interface to both conventional and gradient descent-based ptychographic algorithms.
Features such as mixed-state probe, probe position correction, and multislice ptychography make experimental reconstructions practical and robust.
Reconstructions are specified in a declarative format and can be run from a command line, Jupyter notebook, or web interface. 
Multiple computational backends are supported to provide maximum flexibility.
With the JAX computational backend, a six-fold improvement in iteration speed is achieved over a widely used package implemented in MATLAB, fold\_slice/PtychoShelves.
We report reconstruction success for a variety of experimental datasets, and detail the effects of regularization on convergence and reconstruction quality.
The software promises to speed the application and development of ptychographic methods for materials science.

\end{abstract}
\maketitle
\section{Introduction}

Electron ptychography is emerging as an important technique for the advanced characterization of materials.
In scanning transmission electron microscopy (STEM), ptychography has been used to image materials with record-breaking, deep sub-angstrom resolution \citep{jiang_electron_2018,chen_mixed-state_2020,chen_electron_2021,sha_deep_2022,sha_ptychographic_2023}.
By solving the inverse scattering problem directly, ptychography can account for microscope aberrations and return the specimen potential directly, providing high-contrast information from both heavy and light elements.
Multislice ptychography extends this to 3D, providing depth resolution beyond the diffraction limit \citep{chen_three-dimensional_2021,gilgenbach_three-dimensional_2023,kim_revealing_2024,dong_visualization_2024,zhu_insights_2025}.

As a computational imaging technique, the algorithms that reconstruct ptychographic data are as critical as the experiment that collects the data.
As a result, new developments in ptychography are often preceded and enabled by improvements in the underlying reconstruction algorithms.
The first prominent algorithm for ptychographic reconstruction was Wigner Distribution Deconvolution (WDD) \citep{rodenburg_theory_1992}.
This algorithm deconvolves the 4D dataset into two Wigner functions, one corresponding to the probe and one corresponding to the object. This technique is shown to be relatively robust to noise and to partial coherence.
However, it proved difficult to implement experimentally due to the large quantities of data required and limited processing power at the time \citep{rodenburg_ptychography_2008,rodenburg_ptychography_2019}.

The development of iterative algorithms for ptychography such as the `ptychographic iterative engine' (PIE) \citep{rodenburg_phase_2004} and `enhanced PIE' (ePIE) \citep{maiden_improved_2009} was a key breakthrough, owing to the flexibility and simplicity of such algorithms \citep{rodenburg_ptychography_2019}.
These algorithms proceed probe position-by-probe position, applying a forward simulation of scattering and then updating estimates of the object function using the mismatch between the forward simulated and experimental data. 
Owing to its versatility, the ePIE algorithm has been widely applied to ptychographic reconstruction of X-ray \citep{rodenburg_hard-x-ray_2007}, SEM \citep{humphry_ptychographic_2012}, and STEM \citep{dalfonso_deterministic_2014,wang_electron_2017,jiang_electron_2018} data.

With iterative algorithms, more physically accurate forward models achieve better reconstructions.
For instance, the ePIE algorithm relies on the strong phase object approximation (SPOA) forward model, which breaks down for samples thicker than a few nm in the electron microscope. Multislice ptychography \citep{maiden_ptychographic_2012,tsai_x-ray_2016} replaces this forward model with the multislice model \citep{cowley_scattering_1957} to account for dynamical scattering and thus extend reconstructions to thicker samples.
Mixed state ptychography is another improvement \citep{thibault_reconstructing_2013}, modeling the probe and/or object as a weighted sum of mutually incoherent waves.
In STEM, a mixed state probe is critical to account for partial spatial coherence \citep{chen_mixed-state_2020}, which is on the order of the electron probe's diffraction limit ($\sim\! 0.5 \:\AA$) \citep{dwyer_measurement_2010}.

Beyond improved forward models, advanced gradient update methods can be used, such as the maximum-likelihood algorithm \citep{thibault_maximum-likelihood_2012,odstrcil_iterative_2018}.
Additionally, the choice of loss function allows the noise statistics of the experiment to be incorporated \citep{godard_noise_2012,thibault_maximum-likelihood_2012}.
Alternatively, the iterative problem can be framed as a constraint satisfaction problem, and solved with algorithms such as the difference map \citep{giewekemeyer_quantitative_2010,thibault_probe_2009}. 
Together, these algorithmic improvements have enabled deep sub-angstrom resolution in electron ptychography \citep{chen_mixed-state_2020,chen_electron_2021}.

Recently, there has been a focus on ptychographic reconstructions performed with gradient descent (sometimes referred to as ``Wirtinger flow'') and often implemented via autodifferentiation software \citep{van_den_broek_general_2013,candes_phase_2015,nashed_distributed_2017,kandel_using_2019,schloz_overcoming_2020,du_adorym_2021,seifert_efficient_2021,diederichs_exact_2024}.
Framing the ptychographic problem this way allows easy implementation of new forward models and reconstruction variables, as the gradient step can be calculated automatically from the forward model.

These advancements have spawned a wide range of software packages for ptychography \citep{enders_computational_2016,wakonig_ptychoshelves_2020,du_adorym_2021,loetgering_ptylabmpyjl_2023,diederichs_exact_2024,lee_ptyrad_2025}.
However, there is still a need for software which acts as a platform for developing and applying next-generation algorithms for multislice electron ptychography, without compromising performance or ease-of-use by users at all experience and skill levels.

Here, we present \code{phaser}, an open-source python package for performing ptychographic reconstructions. 
\code{phaser} provides a unified interface for both conventional ptychography algorithms (ePIE and LSQML) as well as gradient descent-based algorithms. 
The forward model incorporates multislice, multiple incoherent probe modes, position correction, and propagator tilt, enabling an accurate match to experiment.
A modular architecture provides flexibility in the reconstruction process while retaining speed and ease of use.
A client-server architecture enables scaling computation from a single computer to clusters and high performance computing, and provides immediate visualization of reconstruction results. 

\begin{figure*}[t]
\includegraphics[width=5in]{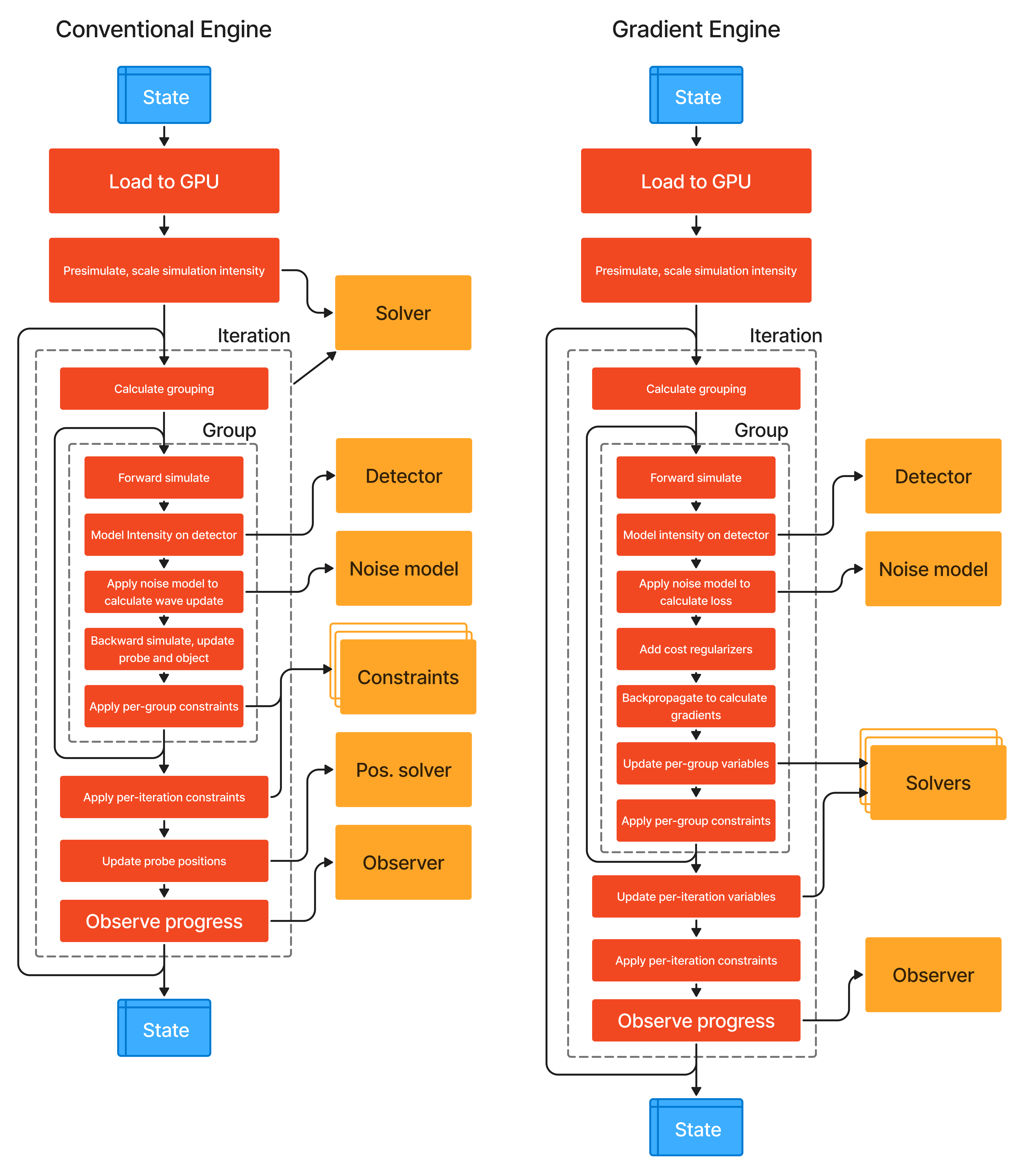}%
\caption{
\label{fig:engines} \textbf{Flowchart of reconstruction engines.}
Flowcharts describing the (a) conventional and (b) gradient descent engine algorithm.
Both engines begin by loading a stored state to the GPU and presimulating the expected diffraction intensity (red blocks).
Each reconstruction iteration (black dashed frame) is subdivided into groups of diffraction patterns which are processed in parallel.
Several components are implemented through modular hooks (yellow boxes), which can be configured by the end user to customize functionality.
}
\end{figure*}

\begin{figure*}
\includegraphics[width=6.2in]{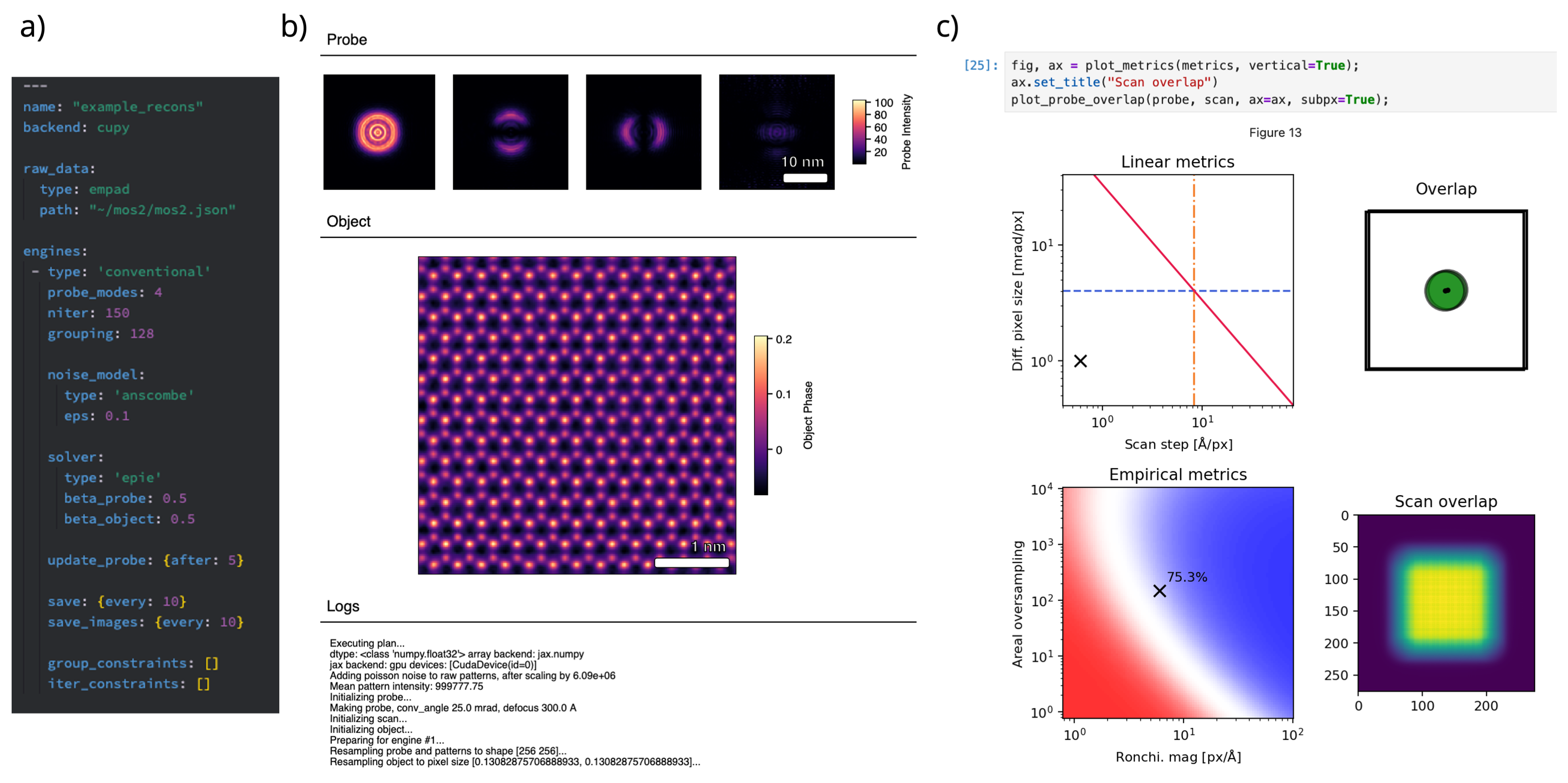}%
\caption{
\label{fig:interface}
\textbf{User interface of \code{phaser}.}
a) Example reconstruction plan file for single slice ptychography with the ePIE engine.
Reconstruction plans are specified as declarative YAML files.
b) Web interface allowing for remote job submission and live viewing of reconstruction process. Probe and object phase are visible as reconstruction proceeds, as well as log messages and errors from the reconstruction.
c) Notebook interface, which provides utilities for viewing raw reconstruction data and analyzing its quality, as well as providing an interface to perform reconstructions.
The plots shown display the dataset's acquisition parameters with reference to the fundamental ptychographic sampling \citep{edo_sampling_2013}, linear oversampling, probe sampling, areal oversampling, and Ronchigram magnification \citep{gilgenbach_methodology_2024}, and provide a view of probe overlap for two scan positions and the entire scan \citep{skoupy_newcomers_2024}.
}
\end{figure*}

\section{Software architecture}

\code{phaser} is designed with a modular architecture, which allows highly configurable reconstructions through a common interface.
Reconstructions in \code{phaser} are specified through `reconstruction plans', which are a declarative description of the reconstruction process to perform, stored as YAML or JSON markup files.
These plans specify data-loading and preprocessing options, followed by a sequence of `reconstruction engines', which are called in sequence to perform reconstructions.
Multiple engines specified in sequence allow the user to progressively regulate the optimization problem, allowing for faster reconstructions and better convergence.

\subsection{Flexibility through modularity}

A key challenge in software design is the balance between the flexibility of the interface, the ease of coding, and the resulting ease of use. 
\code{phaser} attempts to overcome this compromise through the use of `hooks', which allow the specification of reconstruction plans in a modular format.
In software design, a `hook' is a point in a system which allows a user to inject code or behaviors.
Hooks have been widely applied in functional and object-oriented frameworks, with the Emacs editor as an early example \citep{stallman_emacs_1981,froehlich_hooking_1997}.
In \code{phaser}, hooks take the form of functions that implement an abstract interface.
The hooks are called by the main reconstruction algorithm to provide extensible points of customization.
Hooks can be specified in the reconstruction plan file and can take parameters specified by the user as well as arguments passed by the software.
Further, hooks may be provided with the default distribution or can be coded by an advanced user.

As an example, a hook is used to implement the detector noise model as part of the reconstruction process.
The end user can choose one of the built-in noise models (Gaussian, Poisson, amplitude, etc.), or provide their custom noise model as a Python function.
At the same time, reconstruction engines use a common interface for noise models, so the choice of noise model is independent from the choice of engine or forward model.

The use of hooks thus allows the core software package to remain simple while retaining flexibility for the advanced user.
Because hooks are configured through the reconstruction plan file, even novice users can take advantage of the configurability they provide.

\subsection{Plan execution}

A reconstruction begins with a reconstruction plan file, which is parsed and validated to produce a reconstruction plan in-memory.
This gives the end user quick feedback on invalid parameters.
Since plans are specified in a declarative format, they may be generated by other code (such as a hyperoptimization framework or a microscope macro) and easily transmitted over the web to remote workers.
The reconstruction plan schema is versioned, and allows the specification of hooks both built-in and third-party.
These features help make ptychographic reconstructions more reproducible in scientific publications.

Next, the reconstruction plan is executed.
This consists of loading the raw data from a hook (handling the specifics of each detector's file format).
Optionally, the raw data hook may supply metadata about the probe and scan (e.g. a focused probe and raster scan).
This information, optionally with the previous reconstruction state, is used to produce two outputs: a set of loaded diffraction patterns and an initial reconstruction state.
These outputs are then passed through  each reconstruction engine in sequence.

After data loading, each reconstruction engine is executed in turn.
Prior to each engine, the reconstruction state is resampled to conform with the engine reconstruction parameters. This includes padding, cropping, and resampling the probe, diffraction patterns, and object, as well as resampling the object slices along the sample thickness and increasing/decreasing the number of probe modes.
The flow of the two main engine types implemented in \code{phaser}---the conventional engine and the gradient descent engine---are indicated in \Cref{fig:engines}.
Both engines begin by loading the data to the GPU, followed by a precalculation step.
This is used to initialize variables needed in the reconstruction, such as the object magnitude and probe magnitude, as well as to ensure that probe intensity matches with the intensity of the experimental data.
Each engine operates on a subset (called a `group') of probe positions in parallel.
This group may be chosen from a sparse or a compact grouping of the overall probe positions. 
By default, this grouping is randomized per iteration of the solver, and the order in which groups are updated is shuffled in each iteration.
This is in accordance with standard practice in deep learning \citep{meng_convergence_2019}.

In either engine, the first step is to run the `forward model', which uses the current probe and object to simulate diffraction patterns for the group of probe positions.
To perform this forward simulation, sub-regions of the object are sampled to form the transmission functions for each probe position.
Each probe mode is independently propagated through these transmission functions, resampled onto the detector, and incoherent modes are summed to produce a set of simulated diffraction patterns.

In the conventional engines, the simulated wavefront, simulated patterns, and experimental patterns are used to calculate a wavefront update, which we term $\chi$.
At each slice in the object, this wavefront update is used to calculate an object update, as well as an update to the entry wavefront at that slice. This entry wavefront update is propagated backwards to become the wavefront update of the previous slice.
Finally, at the entry plane of the object, the probe and probe positions are updated using the final wavefront update.
As the wavefront update is modified at every slice, the probe and object updates no longer point in the direction of steepest descent.

In the gradient descent update, after the signal is resampled onto the detector plane, the noise model is applied to calculate a loss function which represents the number of electrons that are misplaced on the detector.
Added to this loss function are extra regularizer terms, which act to stabilize the object and probe.
Auto-differentiation, implemented by JAX \citep{bradbury_jax_2018}, is used to determine the gradient with respect to each variable of the simulation.
Steps in each of the reconstruction variables are taken as specified by solvers. Hyperparameter schedules (e.g. cosine decay) can be specified for these solvers, allowing the tuning of initial and final convergence rates.

In either engine, after all groups are processed, iteration-level variables are updated and constraints to the solution are applied.
Finally, the current state of the reconstruction is sent to `observers', which monitor the progress of the reconstruction (e.g. printing log information, saving outputs, reporting to the hyperoptimization framework or webserver, or determining convergence).
When finished, each engine returns an updated reconstruction state which is passed to the next engine in a reconstruction plan.

To provide compatibility across a range of operating systems, accelerators, and environments, \code{phaser} supports multiple computational backends.
At the time of publication, the supported backends are \code{numpy} \citep{harris_array_2020}, \code{cupy} \citep{okuta_cupy_2017}, and JAX \citep{bradbury_jax_2018}.
A set of common abstractions are implemented with each of these libraries, and tests ensure identical behavior.
This allows \code{phaser} to operate on Windows, MacOS, and Linux, with CPU, GPU, and tensor-processing unit (TPU) accelerators.
\code{phaser} supports both single- and double-precision calculations, defaulting to single precision. 

In the future, reduced-precision calculations (e.g. 8- or 16-bit floating point numbers) will be explored to provide additional speed improvements, especially in the initial stages of reconstruction.
This has been demonstrated in the training and inference of neural networks, where reduced precision reduces both compute and memory bandwidth requirements \citep{park_training_2018,kuzmin_fp8_2024}. Additionally, next-generation accelerators developed for artificial intelligence may prioritize lower precision computations \citep{venkataramani_rapid_2021}.

\begin{figure}
\includegraphics[width=3.2in]{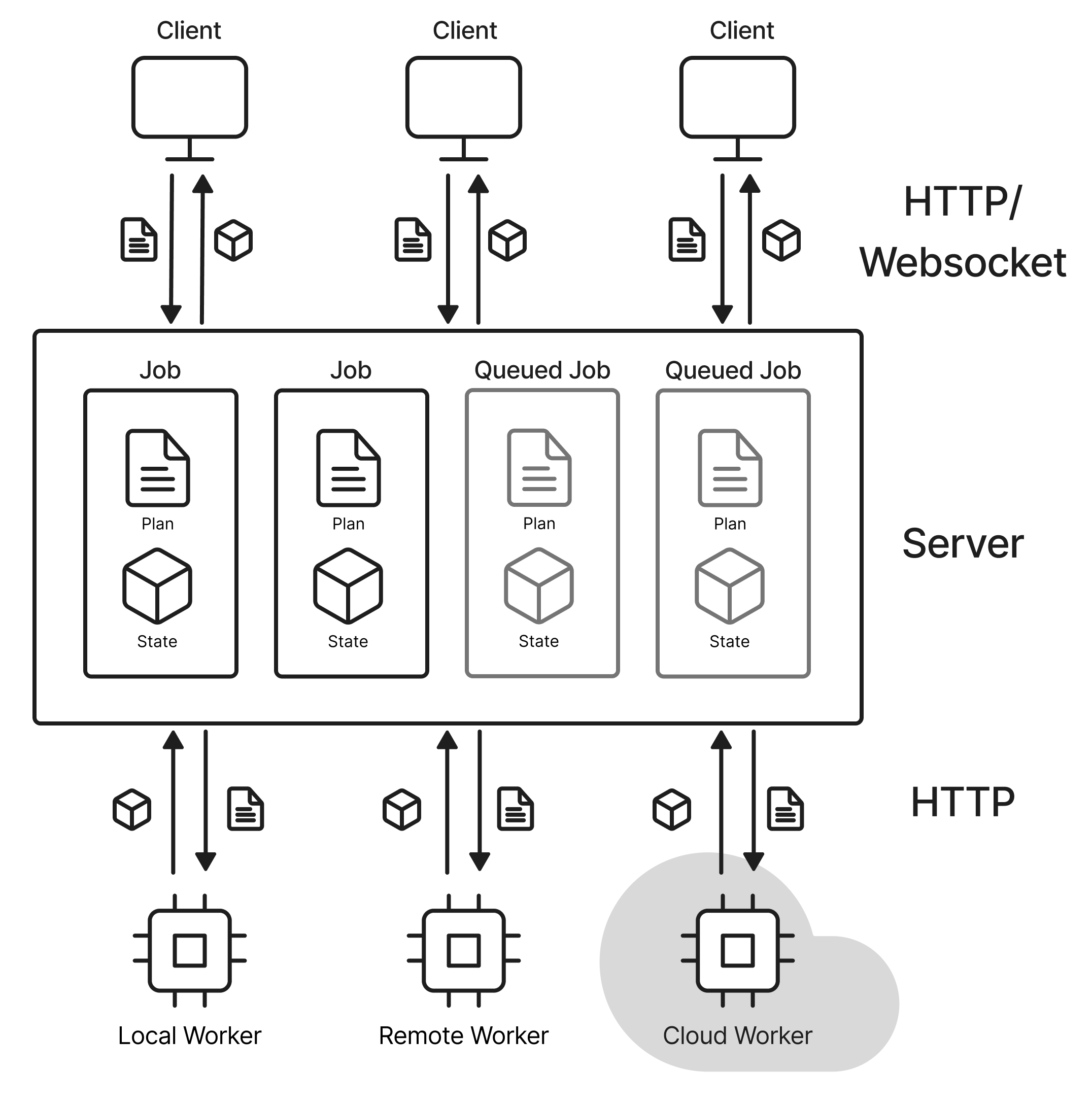}%
\caption{
\label{fig:server-arch}
\textbf{Diagram of server architecture.}
At the center is a server, which serves as a job queue to schedule jobs given by clients (top) to workers (bottom), located locally, remotely, or on the cloud. The reconstruction state is returned by workers to the server, which caches the updated data and publishes to client subscribers. Client-server communications take place over HTTP/Websocket transport, while server-worker communications take place over HTTP with polling.
}%
\end{figure}

\subsection{User interface}

\code{phaser} supports being run from the command line, from an HTML web interface, or through a Jupyter notebook interface as outlined in \cref{fig:interface}.

In each case, the user begins with a reconstruction plan file (\cref{fig:interface}a), specified in a YAML or JSON format.
To run from the command line, this plan file is passed to the \code{phaser run} sub-command, which performs a single reconstruction in batch mode.
To run from the web interface, first a server is started with the \code{phaser serve} sub-command, either locally or remotely.
Then, using a web browser, the user selects a reconstruction plan file and schedules a reconstruction job, and may track its progress.

Run from a Jupyter notebook environment, the \code{phaser} server and the client interface appear as widgets in the Jupyter environment, and can be run locally or remotely.
Jobs can be scheduled from the notebook to worker processes, and updates can be handled by custom code in the notebook.
In addition, \code{phaser} implements several Jupyter widgets which allow visualization of raw data, acquisition parameters, and reconstructed state (\cref{fig:interface}c).
For example, widgets are implemented to show individual convergent-beam electron diffraction (CBED) patterns, position-averaged CBED (PACBED) patterns \citep{lebeau_position_2010}, and virtual detector images.
Widgets also allow the visualization of important acquisition parameters in ptychography, including the fundamental ptychographic sampling \citep{edo_sampling_2013}, probe overlap in a single scan position and summed across the scan \citep{skoupy_newcomers_2024}, probe sampling, and the Ronchigram magnification \citep{gilgenbach_methodology_2024}.
Reconstructed probes can be viewed in real and reciprocal space, and reconstructed objects can be viewed in two and three dimensions.

Underlying the web and notebook interfaces is a general worker-server architecture connecting clients to reconstruction jobs running on workers (\cref{fig:server-arch}).
Workers can run in a separate process on the local computer, run on a remote computer, run on a supercomputing environment, or in the cloud.
Workers poll the server over HTTP to receive jobs to run.

The server has two main functions. The first is to act as a job queue; clients append jobs to this queue to schedule reconstructions, while workers poll the server to request jobs from the queue.
The second main function of the server is as a publisher/subscriber (pub/sub) server; workers communicate updates on the reconstruction to the server, which caches the current reconstruction state in memory and distributes updates to clients, allowing the live viewing of reconstructions.
This feature is critical for previewing the quality of data during a microscope session, and for observing when and why reconstructions begin to diverge.
Communication between the workers and the server takes place over HTTP. Communication between the client and the servers takes place over HTTP/WebSocket transport, which allows for live two-way communication.

\section{Algorithm description}

Ptychography is an inverse problem where, given a series of diffraction patterns taken under shifted illumination, the experimental conditions are found that are `most likely' to yield those diffraction patterns.
These diffraction patterns are most often collected on a pixelated camera, but can also be collected with a segmented detector \cite{zhang_super-resolution_2025}.

Three critical variables contribute to the diffraction patterns measured in ptychography:

\begin{itemize}
 \item The `probe' $P$, a complex field variable indicating the amplitude and phase of the incident wavefunction. In the case of mixed-state ptychography, a set of mutually incoherent `probe modes' $P_k$ is used.
 \item The `object' $O$, which imparts a phase shift and attenuation (amplitude) to the incident probe. The object can be two dimensional (single slice) or three dimensional, in which case slices are separated by gaps of thickness $\Delta z_i$. In electron microscopy, the object slices are sometimes referred to as ``transmission functions''.
 \item Probe positions $X_j$. For each position, the probe is shifted to that location and a diffraction pattern taken.
\end{itemize}

The relevant forward model is the multislice method, which is capable of modeling interaction with a thick specimen as well as multiple scattering. Given a set of probe modes $P_k$, a 3D object $O_i$, and a probe position $P_j$, the wavefunction at each slice is iteratively calculated given the wavefunction at the slice before:
\begin{align*}
  \Psi_{0,k}(\vec{r}) &= P_k(\vec{r} - X_j) \\
  \Psi_{i,k}(\vec{r}) &= (\Psi_{i-1,k}(\vec{r}) \cdot O_{i-1}(\vec{r})) * p(\Delta z_{i-1})
\end{align*}

In the above, $p(\Delta z)$ indicates a Fresnel free-space propagation kernel of distance $\Delta z$, and $*$ is the convolution operator.
A bandwidth limit is also applied at this step to prevent aliasing in frequency space \citep{kirkland_advanced_2010}. This bandwidth limit is customizable, which is particularly useful for unpadded reconstructions where the Nyquist limit can be relatively small in comparison to traditional multislice simulations.

Finally, given the exit wavefunction $\Psi_{n,k}$, the final intensity in reciprocal space may be calculated:

$$
I(\vec{k}) = \sum_{k} \left| \mathcal{F}(\Psi_{n, k})(\vec{k}) \right|^2
$$

The inverse problem consists of taking measured diffraction patterns $I_{exp}$ and recovering the probe $P$ and object $O$.
This inverse problem is known to be unambiguous under certain conditions. For instance, in single-slice ptychography, with known probe positions (which do not fall onto a perfect raster grid), the solution is unambiguous up to a scaling factor of intensity and an affine phase ramp of the object \citep{fannjiang_raster_2019}.
In practice, however, the probe positions $P_j$ are not known perfectly, and initial estimates are updated as the algorithm proceeds. This can introduce ambiguity; in the geometrical optics limit, a change in first order aberrations is equivalent to a linear transformation of the probe positions.

\subsection{Noise models}\label{sec:noise}

Ptychography is an overdetermined nonlinear inverse problem.
Because the problem is overdetermined, the vast majority of experimental datasets have no exact solution; any experimental noise whatsoever will almost certainty perturb the problem into this region.
This problem is overcome by the use of maximum likelihood estimation; rather than attempting to find an exact solution, a solution is found which is `most likely' to generate the recorded data given some model of the experimental noise.

The choice of noise model for reconstruction has been well-covered in the literature \cite{godard_noise_2012,odstrcil_iterative_2018,leidl_influence_2024}. Given a modeled intensity $I(\vec{k})$ and a measured intensity $I_{exp}(\vec{k})$, the ideal solution is one which maximizes the probability $P(I | I_{exp})$, i.e the most likely intensity given the experimental data. This is known as the maximum \textit{a posteriori} estimate, which can be obtained using Bayes' theorem:
\begin{equation*}
P(I | I_{exp}) = \frac{P(I_{exp} | I) P(I)}{P(I_{exp})}
\end{equation*}
Absent an estimate of the prior probabilities $P(I)$, a uniform prior distribution of $P(I)$ (which maximizes $P(I | I_{exp})$) is equivalent to maximizing the `likelihood' $P(I_{exp} | I)$:
\begin{equation*}
\max_{I} P(I_{exp} | I) = \max_{I} \prod_{\vec{k}} P(I(\vec{k}) | I_{exp}(\vec{k}))
\end{equation*}

For Gaussian noise of variance $\sigma^2$ this likelihood is:
\begin{align*}
P(I | I_{exp}) &= \prod_{\vec{k}} \frac{1}{\sigma \sqrt{2 \pi}} \exp \frac{-(I(\vec{k}) - I_{exp}(\vec{k}))^2}{2 \sigma^2} \\
\mathcal{L}(I) &= - \log P(I | I_{exp}) \\
\mathcal{L}(I) &= \sum_{\vec{k}} \frac{1}{2 \sigma^2} \left( I(\vec{k}) - I_{exp}(\vec{k}) \right)^2 + \log \left( \sigma \sqrt{2 \pi} \right)
\end{align*}
As is customary, the loss function $\mathcal{L}(I)$ is defined as the negative log-likelihood. The second term above is a normalization constant and can be ignored. 

With Poisson noise, variance is not constant, but scales with mean intensity.
Therefore, least-squares error cannot be used as an estimator.
Instead, the likelihood is:
\begin{gather*}
P(I | I_{exp}) = \prod_{\vec{k}} \frac{I(\vec{k})^{I_{exp}(\vec{k})} e^{-I(\vec{k})}}{I_{exp}(\vec{k})!} \\
\mathcal{L}(I) = \sum_{\vec{k}} I(\vec{k}) - I_{exp}(\vec{k}) \log I(\vec{k}) + \log\left(I_{exp}(\vec{k})!\right) \\
\mathcal{L}(I) \approx \sum_{\vec{k}} I(\vec{k}) - I_{exp}(\vec{k}) \left( \log I(\vec{k}) - \log I_{exp}(\vec{k}) + 1\right),
\end{gather*}
where at the last step Stirling's approximation has been applied.

In practice, a small offset $\epsilon$ must be added to prevent divergences inside the logarithms.
The epsilon value can be rationalized as a minimum signal recognizable by the detector. As such, signals significantly below this value are assumed to be corrupted by Gaussian noise.
When electron counts are moderate, a variance stabilizing transform can be applied, which transforms Poisson distributed data to an approximately Gaussian distribution, allowing the use of a least-squares estimator.
This leads to the amplitude and Anscombe noise models.
Given a transformation $x \mapsto 2 \sqrt{x + c}$, the transformed variable can be modeled as Gaussian with unit variance. Once transformed, the Gaussian likelihood can be used:
\begin{align*}
\mathcal{L}(I) = \sum_{\vec{k}} \frac{1}{2} \left( \sqrt{I(\vec{k}) + c} - \sqrt{I_{exp}(\vec{k}) + c} \right)^2,
\end{align*}
where $c = 0$ leads to the amplitude noise model and $c = 3/8$ leads to the Anscombe noise model.
The amplitude and Anscombe noise models have the benefit that additive Gaussian noise can be considered analytically, as discussed by \citet{godard_noise_2012}.

For the conventional engines, gradients of the loss functions are taken analytically:
\begin{align*}
\nabla \mathcal{L}_{p}(\Psi) &= \left(1 - \frac{I_{exp}(\vec{k})}{\epsilon + I(\vec{k})} \right) \Psi(\vec{k}) \\
\nabla \mathcal{L}_{a}(\Psi) &= \left(1 - \frac{\sqrt{I_{exp}(\vec{k}) + c}}{\epsilon + \sqrt{I(\vec{k}) + c}} \right) \Psi(\vec{k}),
\end{align*}
where $\Psi(\vec{k})$ is the complex wavefunction on the detector plane.
As noted by \citet{leidl_influence_2024}, these two gradients show significant differences in their frequency spectrum, with the Poisson gradient providing the largest updates at large scattering angles where signals are weak.

Using these gradients, an optimal step size can be calculated \citep{odstrcil_iterative_2018} and a total wavefunction update can be found as $\Delta \Psi(\vec{k}) = - \alpha \nabla \mathcal{L}(\Psi)$. Considering the case of the generalized amplitude loss function, the optimal wavefunction update is:
\begin{align*}
\Delta \Psi(\vec{k}) = \frac{\sqrt{I_{exp}(\vec{k}) + c}}{\epsilon + \sqrt{I(\vec{k}) + c}} \Psi(\vec{k}) - \Psi(\vec{k})
\end{align*}
When using the amplitude noise model ($c=0$), this corresponds to the classic modulus constraint of the ePIE method \citep{odstrcil_iterative_2018}.

\subsection{Gradient descent solver}

The gradient descent engine employs traditional machine learning algorithms to fit the system to the experimental data, minimizing the loss function $\mathcal{L}$, which consists of the loss on the detector plus the loss of each cost regularizer (described in Section \ref{sec:reg}).
Autodifferentiation is used to efficiently obtain the local gradient $\nabla \mathcal{L}$ of the loss with respect to each optimization variable (known as the vector-Jacobian product).
Since the loss function $\mathcal{L}$ is a non-constant real function, it is not holomorphic.
However, Wirtinger derivatives can be used \citep{candes_phase_2015} to overcome this challenge. For real functions, the two Wirtinger derivatives are equivalent up to a conjugation:
$$
\overline{\frac{\partial f}{\partial z}} = \frac{\partial f}{\partial \tilde{z}},
$$
and the gradient $\mathbb{R} \to \mathbb{C}$ can be taken as:
$$
\nabla \mathcal{L} = \overline{\frac{\partial \mathcal{L}}{\partial z}}
$$

As the gradient calculated from a subset/group of the probe positions is a noisy estimate of the total gradient (considering all probe positions), update proceeds according to stochastic gradient descent---which has been shown to yield better convergence when training neural networks \citep{qian_impact_2020} and in phase retrieval \citep{mignacco_stochasticity_2021,odstrcil_iterative_2018}.
One of several optimizers may be used to perform the final update step for each variable.
The Optax library \citep{deepmind_deepmind_2020} implements a number of gradient processing and optimization algorithms, including stochastic gradient descent (SGD), Adam and its variants \citep{kingma_adam_2017,dozat_incorporating_2016}.
The limited-memory Broyden–Fletcher–Goldfarb–Shanno (LBFGS) optimization and stochastic gradient descent with Polyak-Ribere step size \citep{loizou_stochastic_2021} algorithms are also of interest.
Optimizer hyperparameters may be scheduled to vary as the reconstruction progresses, using set schedules (e.g cosine decay) or arbitrary Python expressions.
The Adam optimizer with a fixed learning rate provides acceptable reconstructions in most cases.
However, other optimization methods may allow significantly faster convergence or the reconstruction of difficult datasets.

\subsection{Conventional solvers}

Along with gradient descent, \code{phaser} implements two conventional ptychography algorithms, ePIE and LSQML.
In single slice ptychography, ePIE and LSQML may be considered gradient descent methods with a variable step size.
In multislice ptychography, there is a subtle difference; the conventional engines form an estimate of the optimized wavefront $\Psi$ at each step of backpropagation.
This optimized $\Psi$ is used while calculating the gradient of the previous step.
In contrast, the gradient descent engine takes gradients with respect to each object slice simultaneously, and a step is taken in the direction of the overall object gradient.
\subsubsection{ePIE}
In the ePIE algorithm, the noise model is first used to compute a wavefront update $\chi(\vec{k})$ on the detector.
This update is propagated backwards to the exit plane of the sample.
Then, at each slice, the wavefront update is split into an update applied to the object slice $O_i$, and to the previous wavefront/probe $\Psi_i$:
\begin{align*}
\chi_{n}(\vec{r}) &= \mathcal{F}^{-1}(\chi(\vec{k})) \\
\chi_{i-1} &= \frac{O_i^*(\vec{r})}{\max_{\vec{r}} \left| O_i(\vec{r}) \right|^2} \chi_i(\vec{r}) \\
\Delta O_i &= \frac{P_i^*(\vec{r})}{\max_{\vec{r}} \left| \Psi_i(\vec{r}) \right|^2} \chi_i(\vec{r}) \\
\end{align*}
Probe updates are averaged across the group/batch of positions, while object updates are summed across the group (as well as incoherent probe modes):
\begin{align*}
P(\vec{r}) &\mathrel{+}= \beta_{probe} \frac{\sum_{k} \chi_{0,k}(\vec{r})}{N_k} \\
O_i(\vec{r}) &\mathrel{+}= \beta_{object} \sum_{k} \Delta O_{i,k}(\vec{r})
\end{align*}
This multislice generalization of ePIE (sometimes termed 3PIE) was introduced by \citet{maiden_ptychographic_2012} and is further discussed by \citet{tsai_x-ray_2016}.
One can recognize $O^*(\vec{r})$ as the Wirtinger derivative $\frac{\partial}{\partial \tilde{z}}$ of $P O$ with respect to $P$, confirming that single-slice ePIE can be considered a gradient descent method.
\subsubsection{LSQML}
The multislice LSQML algorithm is implemented as described in \citet{odstrcil_iterative_2018} and \citet{tsai_x-ray_2016}.
Starting with the wavefront update in realspace $\chi_{n,k}(\vec{r})$, the illumination and object update directions are first computed using steepest descent \citep[][eq 24]{odstrcil_iterative_2018}:
\begin{align*}
    \Delta P_{i,k}(\vec{r}) &= \chi_{i,k}(\vec{r}) O^*(\vec{r}) \\
    \Delta O_{i,k}(\vec{r}) &= \chi_{i,k}(\vec{r}) P^*(\vec{r})
\end{align*}
The illumination update direction $\Delta P_{i,k}$ is backwards-propagated and serves as the wavefront update $\chi_{i-1,k}$ for the previous slice.
The object update direction is averaged across probe modes and group, and the step size is calculated per-group \citep[][eq 23,25]{odstrcil_iterative_2018}:
\begin{align*}
    \Delta \hat{O}_{i} &= \frac{\sum_{group} \Delta O_{i,k}(\vec{r})}{\sum_{iter,k} \left| P_{i,k}(\vec{r}) \right|^2 + \delta_O} \\
    \alpha_{O,i} &= \frac{\sum_{\vec{r},k} \mathrm{Re}\left[ \chi_{i,k}(\vec{r}) \left( \Delta O_{i,k}(\vec{r}) P_{i,k}(\vec{r}) \right)^* \right]}{\sum_{\vec{r},k} \left| \Delta O_{i,k}(\vec{r}) P_{i,k}(\vec{r}) \right|^2 + \gamma}
\end{align*}
The final slice update is calculated as \citep[][eq 27]{odstrcil_iterative_2018}:
\begin{align*}
O_i(\vec{r}) &\mathrel{+}= \beta_{object} \frac{\sum_{group} \alpha_{O,i} \Delta \hat{O}_{i}(\vec{r}) \sum_{k} \left| P_{i,k}(\vec{r}) \right|^2}{\sum_{group,k} \left| P_{i,k}(\vec{r}) \right|^2}
\end{align*}

At the last slice, the probe update is calculated similarly to the object.
The intensity sums $\sum_{iter} \left| P_{i,k}(\vec{r}) \right|^2$ and $\sum_{iter} \prod_i \left| O_{i,k}(\vec{r}) \right|^2$ are computed progressively, with the final sum from iteration $i-1$ serving as the value for iteration $i$.
The summed object intensity is recorded in the probe/cutout view, and encodes how transparent the object is to a given region of the probe. The summed probe intensity is recorded in the object view, and encodes the degree of illumination at each pixel of the object.

\subsection{Regularization}
\label{sec:reg}

As ptychography is an ill-posed problem, regularizations are required to stabilize the most likely solution, especially in the presence of noise.
Regularizations act to either constrain the solution set to a desired subspace or to penalize unwanted features in the reconstruction, increasing convexity and prioritizing physically realistic solutions.
The most critical regularization is the maximum likelihood noise model (\cref{sec:noise}).
However, in ptychography several other regularization types have been explored.
For example, total variation (TV) regularization of the object phase has been shown to aid reconstructions in low-dose imaging \citep{tanksalvala_nondestructive_2021}.

Several regularizers are built-in to \code{phaser}, and are listed below.
Additional regularizers can easily be implemented through user-defined hooks.
\code{phaser} implements two main types of regularization. The first, known as constraint regularizers, are called per group or per iteration to constrain the reconstruction to some suitable subset of possible reconstruction states.
Constraint regularizers are supported by all reconstruction engines.

Key constraint regularizers implemented by \code{phaser} are listed below:
\begin{description}
    \item[Object amplitude constraint] Limits the object amplitude to within a specified range.
    Avoids the ambiguity noted by \citet{fannjiang_raster_2019} whereby a scaling of the object amplitude can be compensated by a scaling of the illumination (or another object slice).
    \item[Probe support constraint] Limits the probe to a certain support in reciprocal or real space. Support constraints are widely applied to reduce ambiguity in phase retrieval applications  \citep{fienup_phase_1982,elser_phase_2003}
    \item[Layer regularization] Applies a Gaussian blur to the object in the depth axis. This blur is implemented with a real-space convolution to avoid information bleeding from the top of the reconstructed object to the bottom due to the the assumption of periodic boundary conditions when filtering in reciprocal space.
    \item[Object low pass] Applies a Gaussian low pass filter to the object in the plane.
    \item[Object phase deramp] Removes an affine phase ramp from the object, one of the inherent ambiguities of ptychography \citep{fannjiang_raster_2019}.
\end{description}

The second type of regularization, known as cost regularizations, are added onto the detector loss in the gradient descent engine.
These act to penalize physically unrealistic solutions and to suppress noise in the reconstruction.

Key cost regularizers implemented by \code{phaser} are listed below:
\begin{description}
    \item[L2 regularization] (object, probe) L2 regularization (also known as ridge regression) acts to penalize excessive `energy' in a signal; in this case it penalizes a nonzero probe intensity or a non-vacuum object.
    \item[L1 regularization] (object, object phase, object power spectrum) L1 regularization acts similarly to L2 regularization, but tends towards sparsity. This has been applied to the object power spectrum to bias the reconstruction towards periodic solutions \citep{diederichs_exact_2024}.
    \item[Tikhonov regularization] (object, probe phase plate) Tikhonov regularization is applied to a finite difference operator, penalizing high frequency variation in the object or probe phase. This has the tradeoff of blurring the resulting image.
    \item[TV regularization] (object, probe phase plate) Similar to Tikhonov regularization, but enforces sparsity of the spatial derivative. Tends to create blocky, step-like features, but has been shown to improve reconstructions in low dose conditions \citep{tanksalvala_nondestructive_2021}.
    \item[Tikhonov/TV layer regularization] Tikhonov regularization applied in the depth direction of the object. Performs a similar role as the `layers' constraint regularization. TV regularization may be useful in cases where the object may change suddenly in depth (e.g a heterogeneous interface or phase boundary).
\end{description}

\section{Results}

\subsection{Performance benchmarking}

Two factors are important for performance of a ptychography package.
First, reconstruction speed must be benchmarked, including the speed of each iteration as well as the total time to convergence.
Second, the final quality of reconstructions must be measured. Both are analyzed here.

\begin{figure}
\centering
\includegraphics[width=2.8in]{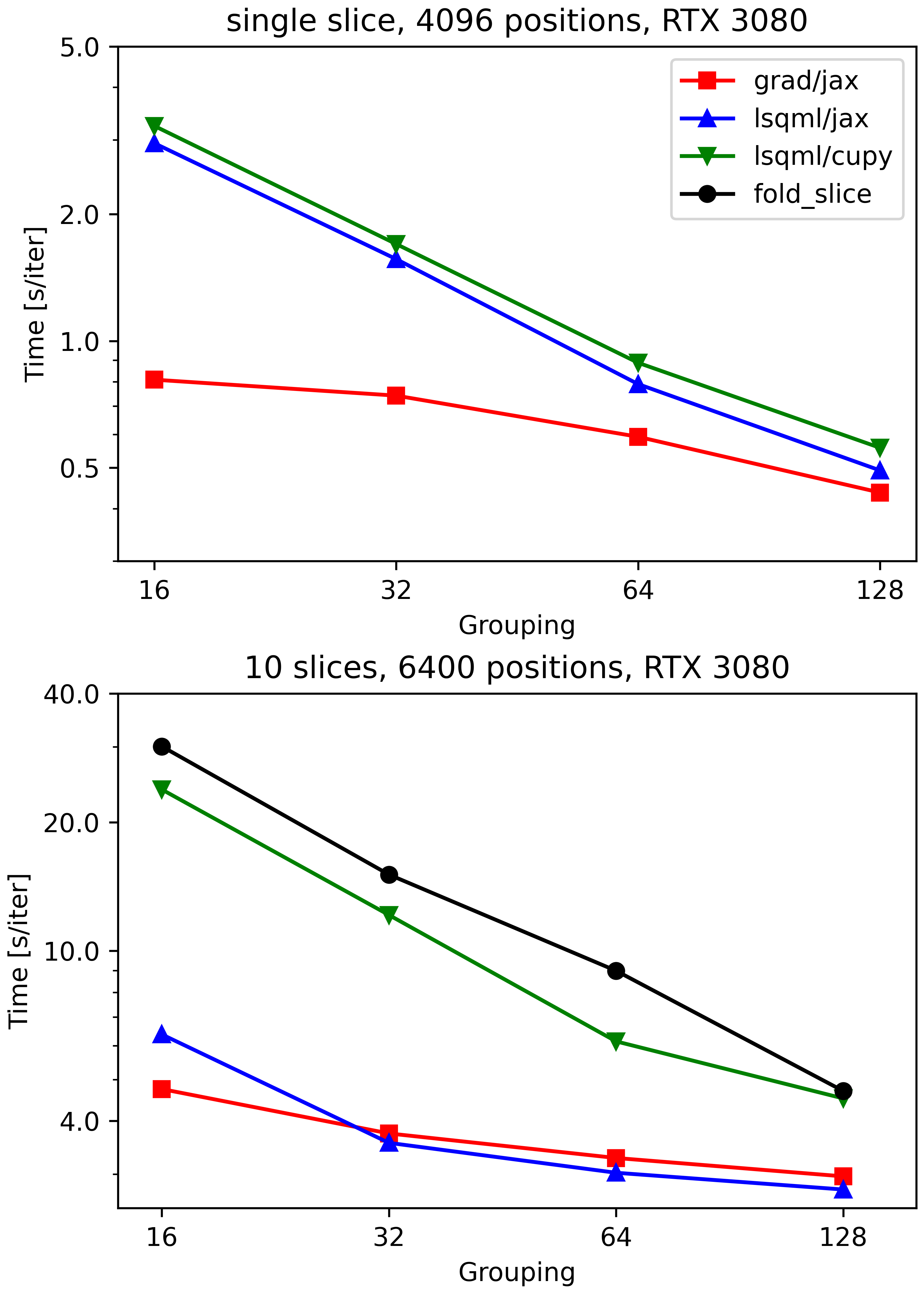}%
\caption{
\label{fig:benchmark-1}
\textbf{Reconstruction performance benchmarking.}
Reconstruction speed (measured as seconds per iteration) for a (top) single-slice and (bottom) multislice dataset of 128x128 diffraction patterns reconstructed with multiple engines and computational backends.
Smaller values indicate faster reconstructions.
}\end{figure}

\begin{figure}
\centering
\includegraphics[width=3.2in]{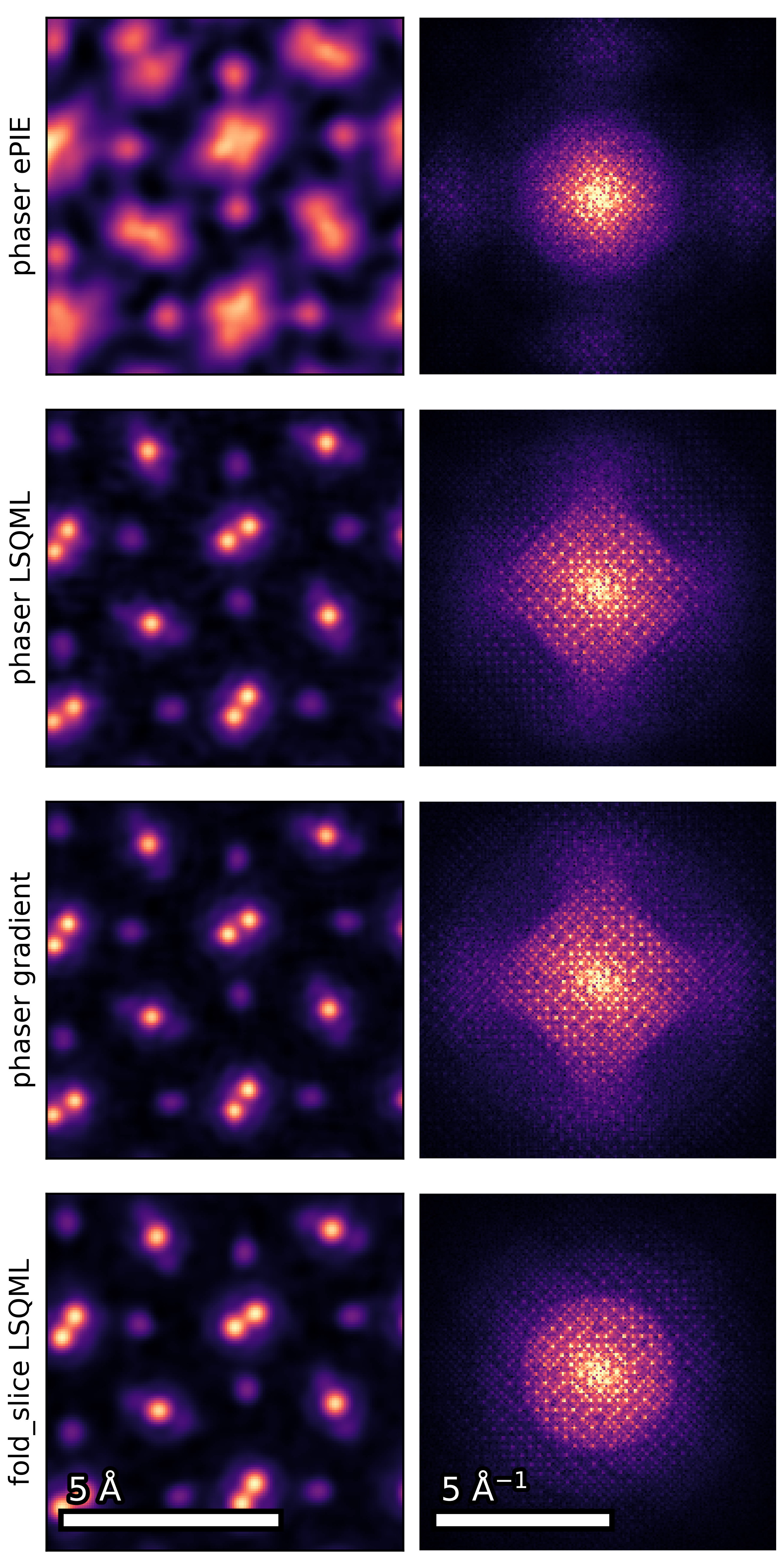}%
\caption{
\label{fig:engine-comparison}
\textbf{Comparison of reconstruction engines}
\ce{PrScO3} dataset \citep{chen_electron_2021} reconstructed with \code{phaser}'s ePIE and gradient descent engine, as well as \code{fold\_slice}'s implementation of LSQML.
The left column displays reconstructed object mean phase, while the right column displays the frequency spectrum (with a gamma of 0.2).
Both LSQML and gradient descent provide superior performance to ePIE.
}
\end{figure}

\Cref{fig:benchmark-1} demonstrates the reconstruction performance of \code{phaser}.
Performance depends heavily on grouping; large groupings are the fastest, at the expense of GPU memory.
At small groupings, more time is spent in the relatively-slow Python interpreter.
Grouping also affects convergence behavior, with larger groupings leading to more averaging of the update steps.
All engines demonstrate improved performance compared to \code{fold\_slice}.

The greatest improvements are found for the JAX backend, which reaches iteration times of less than 3 s/iter for the multislice dataset (6400 probe positions, 20 slices).
This speed is in part due to JAX's just-in-time (JIT) engine, which compiles an optimized GPU kernel for each algorithm.
In ptychography, the bottleneck is the code that runs per group of probe positions, because this code is executed hundreds or thousands of times per iteration.
In \code{phaser} with the JAX backend, this entire inner loop is JIT-compiled, leading to minimal time spent in the relatively slow Python interpreter.
These improvements are most stark at small groupings, where the JAX backend outperforms both the \code{cupy} backend and \code{fold\_slice} by a factor of 5-6x.

\begin{figure}
\centering
\includegraphics[width=2.8in]{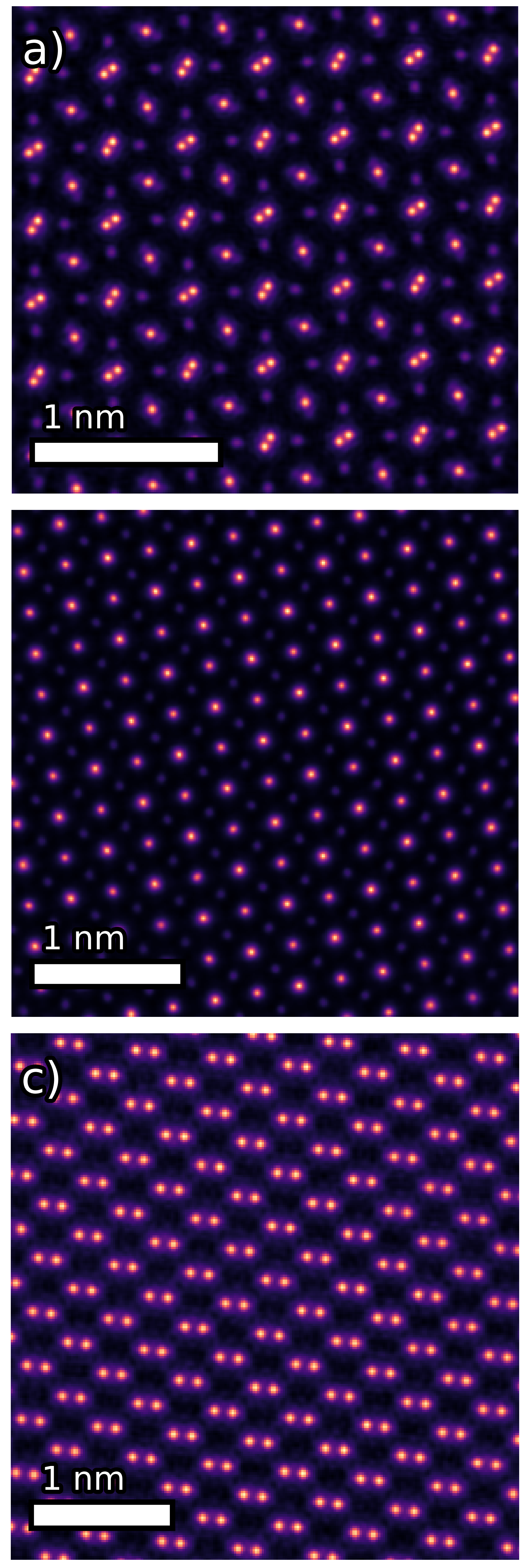}%
\caption{
\label{fig:exp-1}
\textbf{Experimental reconstructions with \code{phaser}}.
Reconstructions were performed with the gradient descent engine.
a) \ce{PrScO3} dataset from Ref.~ \citep{chen_electron_2021}.
b) \ce{BaTiO3} dataset.
c) Si dataset from Ref.~\citep{gilgenbach_methodology_2024}.
}%
\end{figure}

The final quality of reconstructions is shown for the  \ce{PrScO3} dataset from \citet{chen_electron_2021} and reconstructed with different engines in \cref{fig:engine-comparison}.
Both LSQML and gradient descent perform significantly better than ePIE for multislice datasets.
Gradient descent provides slightly better separation of the Pr-Pr dumbbells, as well as information transfer to higher frequencies.
However, both the LSQML and gradient descent reconstructions with \code{phaser} display anisotropic power spectrums.
This may be due to anisotropy in the strength of reflections and therefore signal-to-noise ratio in the sample, or may be an artifact of the object regularizations used (particularly Tikhonov regularization).
Representative experimental reconstructions performed with \code{phaser} are displayed in \cref{fig:exp-1}.
Reconstructions were performed with the gradient descent engine and the Poisson noise model.

\begin{figure*}
\centering
\includegraphics[width=6.4in]{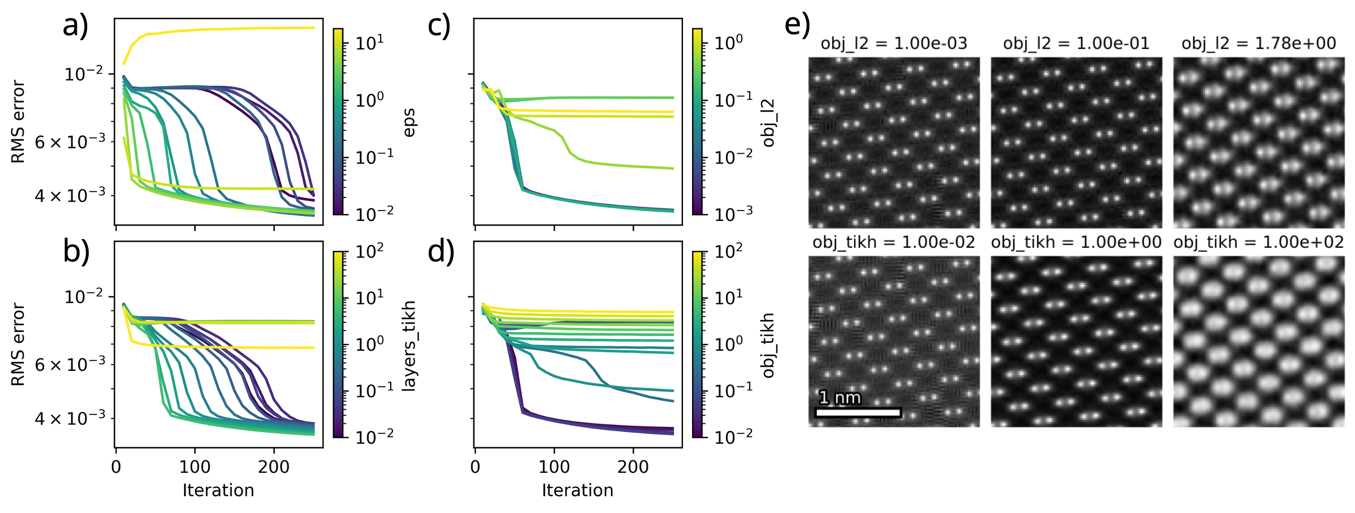}%
\caption{
\label{fig:reg}
\textbf{Impact of regularizations on reconstruction.}
 a-d) Convergence of reconstructions while varying reconstruction parameters / regularizers. a) Poisson noise model epsilon, b) layers Tikhonov regularization, c) object L2 regularization and d) object Tikhonov regularization. Lines display real space RMS error versus reconstruction iteration, with colors indicating regularization value. e) Final reconstructions for selected values of object L2 and object Tikhonov regularizations.
 Errors are shown as root mean squared (RMS) error in real space.
}
\end{figure*}

\subsection{Impact of regularizations}

To understand the impact of the regularization parameters on the reconstruction, a series of reconstructions was performed, varying one parameter at a time. The results are summarized in \Cref{fig:reg}. Empirically, two primary behaviors are observed. Some regularizers, for instance the noise model epsilon and the layers Tikhonov, mainly affect the rate of convergence of the reconstruction. This is seen in fig.~\ref{fig:reg}a and b, where moderate regularization parameter values result in faster convergence. A regularization cost that is too large can, however, prevent convergence entirely. 

Other regularizers show little effect on convergence rate, but still affect final reconstruction quality. This second behavior is observed for the object L2 and object Tikhonov regularizers, as seen in figs.~\ref{fig:reg}c and d. 
\Cref{fig:reg}e shows the final reconstructed object with varying object L2 and object Tikhonov. For object L2, intermediate values lead to the best contrast, while low values result in excessive noise and high values result in degraded resolution. Similar effects are seen for the object Tikhonov.
Fully characterizing the effect of regularizations on reconstruction performance is challenging due to couplings between the regularization parameters.

\begin{figure}
\centering
\includegraphics[width=2.8in]{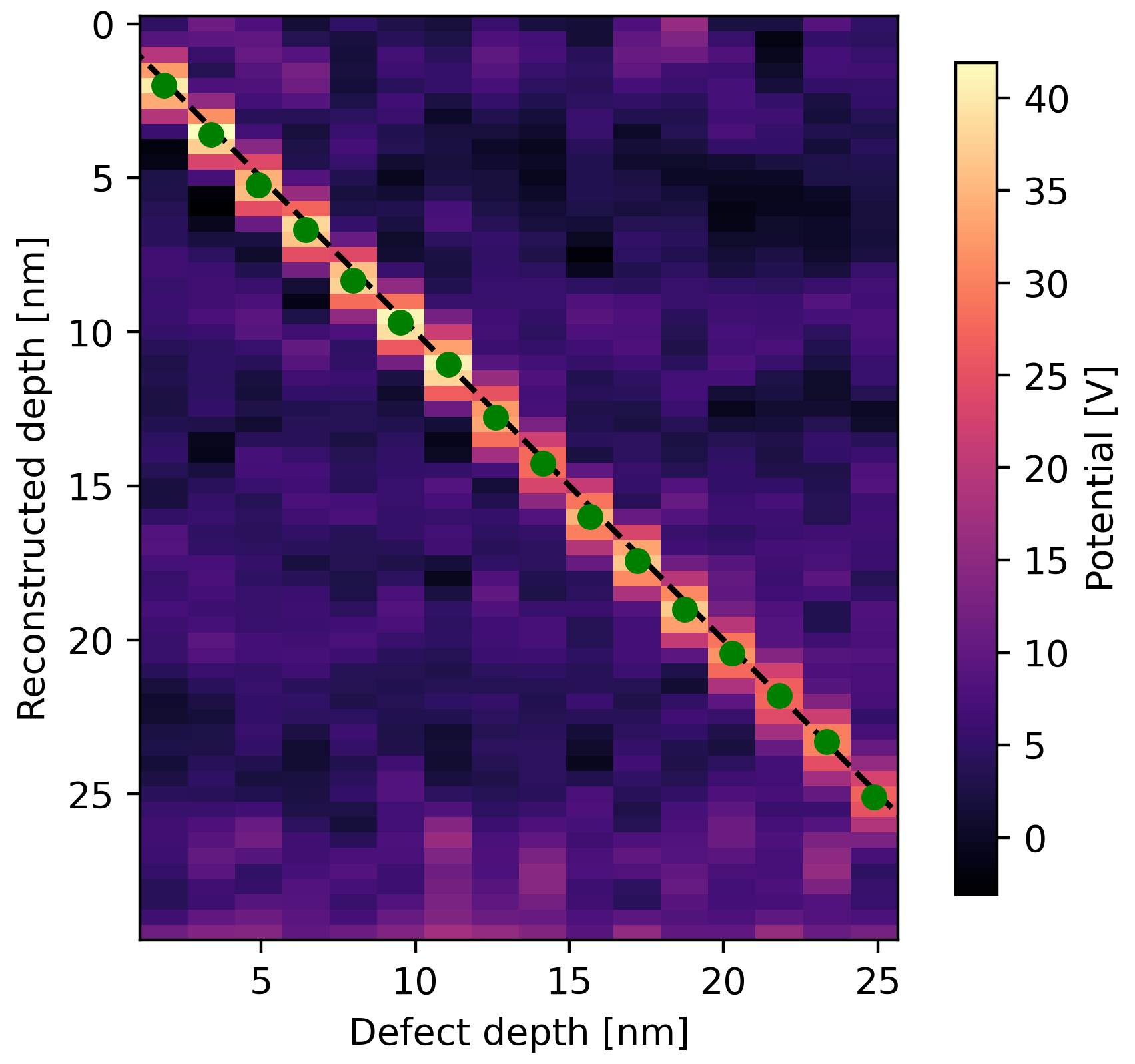}%
\caption{
\label{fig:depth-reg}
\textbf{Depth sensitivity of reconstructions.}
Analysis of reconstruction depth sensitivity for a simulated silicon crystal containing Sn interstitials placed at varying depths. Color represents the reconstructed potential for each interstitial column at each slice in the reconstruction.
The x-axis shows the placed interstitial depth, while the y-axis shows depth in the reconstructed image.
The black dashed line indicates where the reconstructed position equals the actual position, and the green circular symbols represent the reconstructed defect position.
Amorphous carbon surface layers are visible as a diffuse background in the measured potential.
}
\end{figure}

Depth regularization is critical in providing high-quality multislice ptychography reconstructions.
To characterize the depth sensitivity of the gradient descent engine, a reconstruction was performed on a simulated dataset consisting of Sn interstitials in a Si host lattice.
To quantify the precision and resolution in the depth, the reconstructed interstitials are compared to the known interstitial depths in \Cref{fig:depth-reg}.

Reconstructed dopant positions are highly precise, with a root-mean square error in dopant depth of 1.1 $\AA$.
However, depth resolution is limited, with an average full-width half maximum (FWHM) in depth of 1.88 nm. This resolution is best near the top of the sample and worst at the bottom (as can be seen by the gradual blurring of the peaks in \cref{fig:depth-reg}).
As depth resolution scales with the square of numerical aperture, it is expected that higher convergence angles will increase this resolution, provided sufficient dose \citep{chen_three-dimensional_2021}.
As we have demonstrated, the choice of regularizers and other reconstruction parameters can have a strong effect on reconstruction performance and quality.
Furthermore, the couplings between reconstruction parameters make optimization challenging.
To address this challenge, \code{phaser} may be coupled to hyperparameter optimization frameworks, which allow for the optimization of experimental and reconstruction parameters in ptychography \citep{cao_automatic_2022}.
We have coupled \code{phaser} to the Optuna \citep{akiba_optuna_2019} library for hyperoptimization. Optuna implements several optimization algorithms including Bayesian optimization with Gaussian processes (BO-GP) and the tree-structured Parzen estimator (TPE) \citep{bergstra_algorithms_2011}, as well as supporting multi-objective optimization.
A variety of error metrics may be used for optimization. When an object ground truth is available, we have found mean-squared error (MSE) to exhibit better performance than structural similarity (SSIM) \citep{dosselmann_comprehensive_2011,nilsson_understanding_2020}.
Fourier ring correlation (FRC) may be used as well \citep{cao_automatic_2022}.
In the absence of ground truth, the problem is more challenging, as any metric must distinguish between signal and noise, and cannot assume that noise is independently sampled between object pixels. For this reason, the single-image `self-FRC' is often invalid in the setting of ptychography \citep{verbeke_self_2024}.

\section{Conclusions}
\code{phaser} provides a shared, declarative interface to multiple ptychographic reconstructions engines and computational backends---enabling flexibility while retaining ease of use.
This flexibility enables \code{phaser} to be a platform for the future development and application of ptychographic algorithms.
Web, notebook, and command line interfaces are provided, allowing \code{phaser} to scale from a single computer to the cloud and enabling the live viewing of in-progress reconstructions.
With the JAX computational backend, a 6x improvement in iteration speed is achieved over a state-of-the-art package, fold\_slice/PtychoShelves.
\code{phaser} is released under the MPL 2.0 open-source license, and is available on GitHub at \url{https://github.com/hexane360/phaser} and the Python package interface (PyPI) as \code{phaserEM}.

\FloatBarrier
\section{Methods}
Speed was benchmarked using the time per complete iteration as a metric, determined for two representative datasets---one single slice dataset and one multislice with 20 slices.
Each dataset contained 128x128 pixel diffraction patterns. The single slice dataset used a 64x64 scan (4096 total probe positions), while the multislice dataset used a 80x80 scan (6400 positions).
Comparisons with the \code{fold\_slice} fork of PtychoShelves \citep{wakonig_ptychoshelves_2020} used the LSQ-MLs engine, with equivalent reconstruction parameters to the LSQML engine in \code{phaser}.
All benchmarks were performed using a Nvidia RTX 3080 GPU running on an Ubuntu virtual machine.

Quality comparisons were performed on a reference experimental dataset of \ce{PrScO_3} \citep{chen_electron_2021}.
Reconstruction parameters for \code{fold\_slice} were the same as used in \citet{chen_electron_2021}.
For each of the phaser engines, hyperparameter optimization was performed using Optuna.
The error metric used for hyperoptimization was the root-mean squared error of the summed object potential versus a simulated object potential using Kirkland parameterizations \citep{kirkland_advanced_2010} and thermally averaged using isotropic Debye-Waller factors \citep{gesing_refinement_2009}.
Hyperoptimizer parameters and final optimized reconstruction parameters are provided in the Dryad record.

Regularization studies were performed on an experimental Si dataset taken along the $[110]$ zone axis \citep{gilgenbach_methodology_2024}.
Unless otherwise varied, reconstructions used an epsilon of 0.8, object L2 regularization of 0.05, object Tikhonov regularization of 0.05, and a layers Tikhonov regularization of 5.
Again, errors were determined as the root-mean squared error versus a simulated, thermally averaged, object potential.

Depth sensitivity calculations were performed using a Si dataset with 16 Sn interstitials placed at tetrahedral sites.
A 92.3 x 92.2 x 293.5 $\mathrm{\AA}$ supercell was simulated at 300 kV with a convergence angle of 25 mrad, a defocus of 10 nm, and a dose of $1.1 \times 10^7$ $e^-/\AA^2$.
2 nm of amorphous carbon was added to the top and bottom surfaces. 
No lattice strain was incorporated.
Reconstructions were performed with a layers Tikhonov regularization of 10.0.
Interstitial potential was calculated by averaging a circle of 0.2 angstrom radius around each interstitial site as a function of depth.
Finally, interstitial depths and full width half maximum (FWHM) resolution can be found by fitting a 1D Gaussian profile to each interstitial site.
The simulated dataset and final reconstruction are provided in the Dryad record.

\subsection{Code \& Data availability}
\code{phaser} is released under the MPL 2.0 open-source license, and is available on GitHub at \url{https://github.com/hexane360/phaser}.
4D-STEM datasets and reconstruction plan files used in the Results section are available in the Dryad repository [link].
Additional data is available from the authors upon reasonable request.

\section{Acknowledgments}

The authors thank Dr. Jingrui Wei for providing the \ce{BaTiO3} experimental dataset.
CG and JML acknowledge support from the Department of Homeland Security Countering Weapons of Mass Destruction Office (22CWDARI00046-01-00). 
MZ and JML acknowledge support from the Army Research Laboratory, Cooperative Agreement Number W911NF-24-2-0100.
This support does not constitute an express or implied endorsement on the part of the Government.
This work was performed with the assistance of MIT SuperCloud.

\section{Author contributions}

CG and JML conceived of the research and developed the package architecture.
CG and MLZ contributed code to the package.
CG performed benchmarking, experimental reconstructions, and regularization analysis.
All authors read and approved the final manuscript.

\bibliography{references}
\end{document}